\documentclass{article}
\usepackage{spconf,amsmath,graphicx}

\usepackage{enumitem}
\setlist{nosep, leftmargin=14pt}

\usepackage{mwe} % to get dummy images
\usepackage{bm}
\usepackage{amsmath}
\usepackage{amssymb}
\title{Adaptive Whole-Body PET Image Denoising Using 3D Diffusion Models with ControlNet}

\name{Boxiao Yu, Kuang Gong\thanks{For correspondence: kgong@bme.ufl.edu}}
\address{J. Crayton Pruitt Family Department of Biomedical Engineering, University of Florida}
\begin{document}
%\ninept
%
\maketitle
\begin{abstract}
%100-150 words
Positron Emission Tomography (PET) is a vital imaging modality widely used in clinical diagnosis and preclinical research but faces limitations in image resolution and signal-to-noise ratio due to inherent physical degradation factors. Current deep learning–based denoising methods face challenges in adapting to the variability of clinical settings, influenced by factors such as scanner types, tracer choices, dose levels, and acquisition times. In this work, we proposed a novel 3D ControlNet-based denoising method for whole-body PET imaging. We first pre-trained a 3D Denoising Diffusion Probabilistic Model (DDPM) using a large dataset of high-quality normal-dose PET images. Following this, we fine-tuned the model on a smaller set of paired low- and normal-dose PET images, integrating low-dose inputs through a 3D ControlNet architecture, thereby making the model adaptable to denoising tasks in diverse clinical settings. Experimental results based on clinical PET datasets show that the proposed framework outperformed other state-of-the-art PET image denoising methods both in visual quality and quantitative metrics. This plug-and-play approach allows large diffusion models to be fine-tuned and adapted to PET images from diverse acquisition protocols.
\end{abstract}
\section{Introduction}
\label{sec:intro}

Positron Emission Tomography (PET) is a crucial functional imaging modality utilized in both preclinical research and clinical diagnosis. Due to various physical degradation factors, PET images often suffer from low image resolution and signal-to-noise ratio. Performing PET image denoising enhances quantitative accuracy and lesion-detection precision. Given that PET image quality is significantly affected by factors such as scanners, tracers, dose levels, and scanning time, there is an urgent need for PET image denoising methods that can adapt to diverse clinical settings.

With advances in computational power and dataset accessibility, deep learning-based image denoising has been widely studied, achieving remarkable success. Among these, diffusion models~\cite{ho2020denoising} have emerged in recent years as one of the most effective generative models for various image processing tasks. Diffusion models transform data from a normal distribution to the target data distribution through the gradual refinement process. In the forward diffusion process, Gaussian noise is progressively added to the target data until it approximates pure noise. In the reverse diffusion process, noise is incrementally removed to reconstruct the desired target data. Various variants of diffusion models have been successfully applied to PET image denoising tasks~\cite{xie2023ddpet, jiang2023pet, shen2023pet, pan2024full}, demonstrating promising outcomes.

However, existing diffusion model–based denoising methods face challenges in adapting to different acquisition protocols to accommodate diverse PET datasets effectively. For instance, supervised learning-based methods can produce high-quality denoising results, but training a large-scale conditional diffusion model individually for each protocol is impractical and inefficient. Moreover, paired data for some specific protocols are limited in scale. Directly fine-tuning large pre-trained diffusion models with limited data may lead to overfitting and catastrophic forgetting. Zero-shot methods learn only the distribution of high-quality PET images during training and embed low-quality PET images as data-consistency constraints during inference to handle noisy images at various noise levels~\cite{gong2024pet, xie2024dose}. While this approach eliminates the need for repeated training, it lacks the ability to convey fine-grained control over the final generated images, and the denoising results are highly sensitive to the constraint strength. Thus, there is a need for PET image denoising methods that can incorporate target domain-specific information (such as particular PET protocols) in fine-tuning, while preserving the integrity of large pre-trained models.
\begin{figure*}[t]
\includegraphics[width=\textwidth]{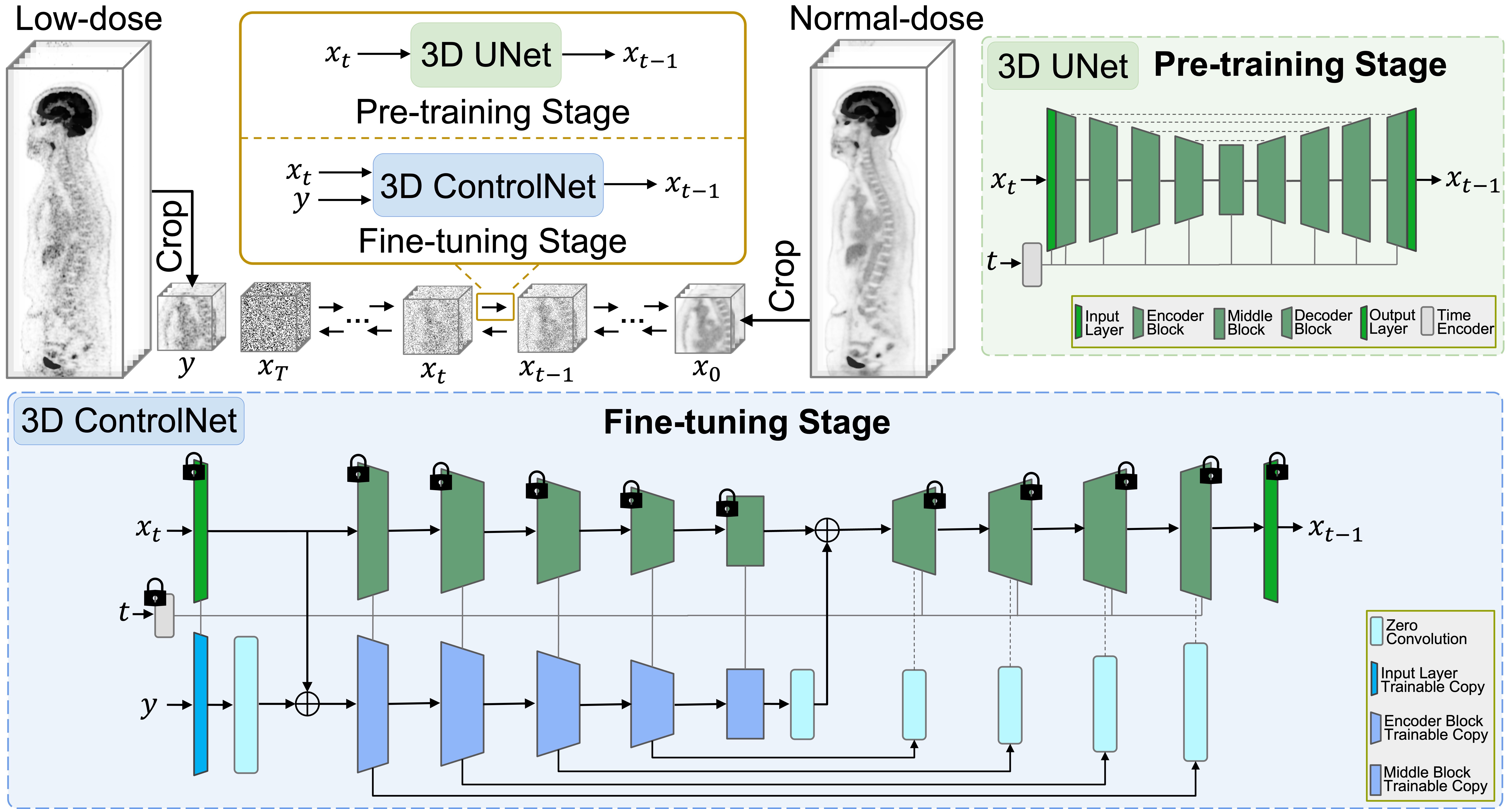}
\caption{Overview of the proposed 3D ControlNet-based denoising model.} \label{fig_model}
\end{figure*}

ControlNet~\cite{zhang2023adding} brings spatial control to large, pre-trained text-to-image diffusion models, and its variants have since been widely applied in image-processing tasks by researchers. However, current ControlNet variants are predominantly based on Latent Diffusion Models (LDMs)~\cite{rombach2022high}, performing denoising on latent representations rather than operating directly in the original PET image space. Although this approach enables more memory-efficient training, it fails to reconstruct the precise 3D local details required for quantitative PET imaging. In this work, we proposed a novel 3D ControlNet-based denoising method that operated directly in the original PET image space to address the limitations of existing approaches. Integrating ControlNet with a pre-trained 3D DDPM makes fine-tuning feasible using a small set of paired low-dose and normal-dose PET images, while preserving the capabilities of the large pre-trained model. This can preserve the precise local details, ensuring high-quality PET images suitable for clinical diagnosis. Experimental results indicate that the proposed approach surpassed existing denoising methods in visual quality and quantitative metrics. This plug-and-play framework is suitable for fine-tuning any large diffusion model operating in the original PET image space, underscoring its potential for denoising PET images across diverse protocols.
\section{Methodology}
\label{sec:method}

\begin{figure*}[t]
\centering
\includegraphics[width=16cm]{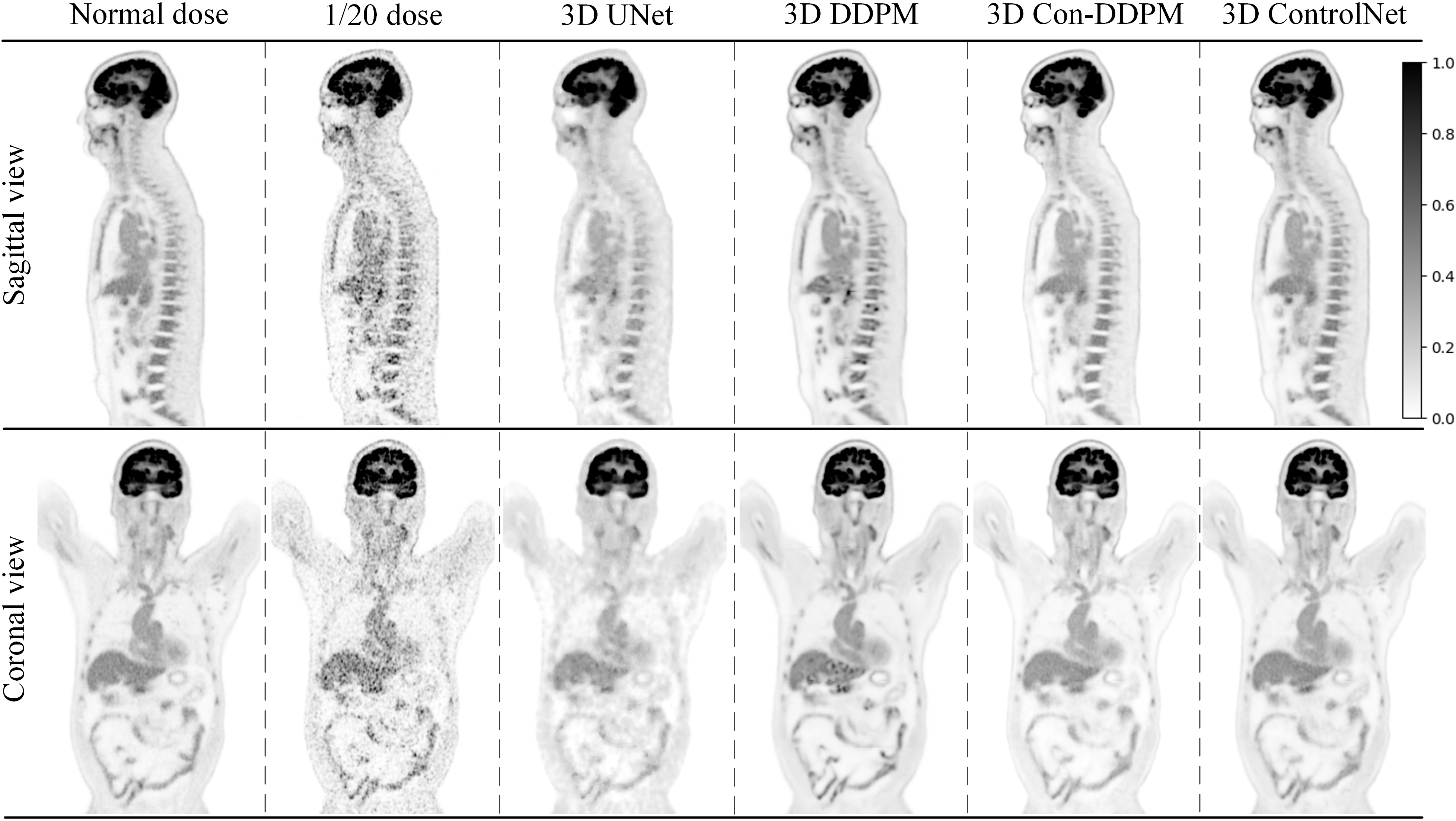}
\caption{Sagittal and coronal view of $1/20$ low-dose PET images and the corresponding denoised results using the proposed 3D ControlNet and other reference methods, along with the normal-dose PET.}
\label{fig_visual}
\end{figure*}

\subsection{Pre-training with 3D DDPM}
In the first stage, as shown in Fig.~\ref{fig_model}, we pre-trained a 3D DDPM~\cite{ho2020denoising} using a large-scale dataset of high-quality normal-dose PET images to effectively learn the complex distribution of PET images. The DDPM operated by progressively adding and then removing noise through a Markov chain of diffusion steps. In the forward diffusion process, Gaussian noise was gradually added to a clean data point $\mathbf{x}_0$ sampled from the real data distribution $q(\mathbf{x})$, following a variance schedule $\{\beta_t \in (0,1)\}^T_{t=1}$ over $T$ time steps
\begin{align} q(\mathbf{x}_t|\mathbf{x}_{t-1}) = \mathcal{N}(\mathbf{x}_t;\sqrt{1-\beta_t}\mathbf{x}_{t-1},\beta_t \mathbf{I}). \end{align}
Using $\alpha_t = 1-\beta_t$ and $\bar\alpha_t = \prod_{s=1}^t \alpha_s$, the forward process could be further expressed as
\begin{align} q(\mathbf{x}_t|\mathbf{x}_0) = \mathcal{N}(\mathbf{x}_t; \sqrt{\bar\alpha_t}\mathbf{x}_0, (1-\bar\alpha_t)\mathbf{I}). 
\end{align}
During the reverse diffusion process, we aimed to reconstruct the clean data from the noisy input by iteratively removing the added noise. Since the true reverse distribution $q(\mathbf{x}_{t-1}|\mathbf{x}_t)$ was intractable, $p_{\boldsymbol{\theta}}$ was trained to approximate it as
\begin{align}  p_{\boldsymbol{\theta}}(\mathbf{x}_{t-1}|\mathbf{x}_t) = \mathcal{N}(\mathbf{x}_{t-1}; \boldsymbol{\mu}_\theta(\mathbf{x}_t, t), \boldsymbol{\Sigma}_\theta(\mathbf{x}_t, t)), \end{align}
where $\boldsymbol{\mu}_{\boldsymbol{\theta}}$ and $\boldsymbol{\Sigma}_{\boldsymbol{\theta}}$ were the predicted mean and variance, parameterized by a neural network with weights $\boldsymbol{\theta}$. The model was then re-parameterized to predict the noise $\boldsymbol{\epsilon}_{\boldsymbol{\theta}}(\mathbf{x}_t, t)$ instead of the mean. Finally, $\mathbf{x}_{t-1}$ could be estimated as
\begin{align} \mathbf{x}_{t-1} = \frac{1}{\sqrt{\alpha_t}} \left( \mathbf{x}_t - \frac{\beta_t}{\sqrt{1 - \bar{\alpha}t}}, \boldsymbol{\epsilon}_{\boldsymbol{\theta}}(\mathbf{x}_t, t) \right) + \sigma_t \mathbf{z}, \end{align}
where $\mathbf{z} \sim \mathcal{N}(0, \mathbf{I})$. The widely used 3D UNet was adopted as the backbone network to train the score function. By leveraging a large-scale dataset of high-quality PET images, the DDPM effectively learned the intricate features and variability inherent in PET imaging. This extensive pre-training enabled the model to generalize well and provided a strong foundation model for subsequent fine-tuning steps. 

\subsection{Fine-tuning with 3D ControlNet}

During the fine-tuning stage, a 3D ControlNet was trained using a small set of paired normal-dose and low-dose PET images. The 3D ControlNet allowed us to incorporate additional low-dose PET image $\mathbf{y}$ into the pre-trained 3D UNet architecture, enabling the 3D DDPM to generate the corresponding normal-dose PET images rather than random samples. By freezing the parameters of the original 3D UNet and creating trainable copies of its encoder blocks, the 3D ControlNet preserved the quality and functionality of the large pre-trained model. The trainable copy and the original model were connected via zero convolution layers (i.e., $1 \times 1$ convolutions with weights and biases initialized to zero, denoted as $\mathcal{Z}(\cdot;\cdot) $), which prevented the fine-tuning process from disruptive interference.

Specifically, in the original 3D UNet, $\mathbf{x}_t$ was first fed into the input layer $\mathcal{F}_I(\cdot; \boldsymbol{\Theta}_I)$ with parameters $\boldsymbol{\Theta}_I$, and then passed through the encoder blocks $\mathcal{F}_E(\cdot; \boldsymbol{\Theta}_E)$ with parameters $\boldsymbol{\Theta}_E$ to obtain the feature map $\mathbf{f}_t$
\begin{align}
\mathbf{f}_t = \mathcal{F}_E\left( \mathcal{F}_I\left( \mathbf{x}_t; \boldsymbol{\Theta}_I \right); \boldsymbol{\Theta}_E \right).
\end{align}
Then, \( \mathbf{f}_t \) was processed by the decoder blocks and output layer to produce the estimated denoised image $\mathbf{x}_{t-1}$.

During fine-tuning, we froze all parameters of the 3D UNet and cloned its input layer and encoder blocks into trainable copy with parameters $\boldsymbol{\Theta}_{IC}$ and  $\boldsymbol{\Theta}_{EC}$. Fig.~\ref{fig_model} illustrates the framework of the 3D ControlNet. The trainable copy took the $\mathbf{y}$ as input, which was passed through the trainable input layer and the following zero convolution layer $\mathcal{Z}_1\left( \cdot; \boldsymbol{\Theta}_{z1} \right)$ with trainable parameters $\boldsymbol{\Theta}_{z1}$. The output was then added with the features from $\mathbf{x}_{t-1}$ to obtain the intermediate feature map  $\mathbf{m}_{tc}$
\begin{align}
\mathbf{m}_{tc} = \mathcal{Z}_1\left( \mathcal{F}_I\left( \mathbf{y}; \boldsymbol{\Theta}_{IC} \right); \boldsymbol{\Theta}_{z1} \right) + \mathcal{F}_I\left( \mathbf{x}_t; \boldsymbol{\Theta}_I \right),
\end{align}
which was then fed into the trainable encoder blocks, followed by another zero convolution layer $\mathcal{Z}_2\left( \cdot; \Theta_{z2} \right)$ with trainable parameters $\boldsymbol{\Theta}_{z2}$. The output was added to the feature map $\mathbf{f}_t$ from the original 3D UNet to obtain the controlled feature map $\mathbf{f}_{tc}$
\begin{align}
\mathbf{f}_{tc} = \mathcal{Z}_2\left( \mathcal{F}_E\left( \mathbf{m}_{tc}; \boldsymbol{\Theta}_{EC} \right); \boldsymbol{\Theta}_{z2} \right) + \mathbf{f}_t.
\end{align}
The estimated denoised image $\mathbf{x}_{t-1}$ was produced by feeding $\mathbf{f}_{tc}$ into freezing decoder blocks and output layer, in which the outputs from the trainable copies of the 3D ControlNet were skip-connected to the original decoder block via zero convolution layers. Our 3D ControlNet performed the denoising process in the PET pixel space, enabling the preservation of precise local details required for quantitative PET imaging.
\section{Experiments and Results}
\label{sec:exp&results}
% 3 page

\subsection{Dataset and Implementation Details}

\begin{figure*}[t]
\centering
\includegraphics[width=16cm]{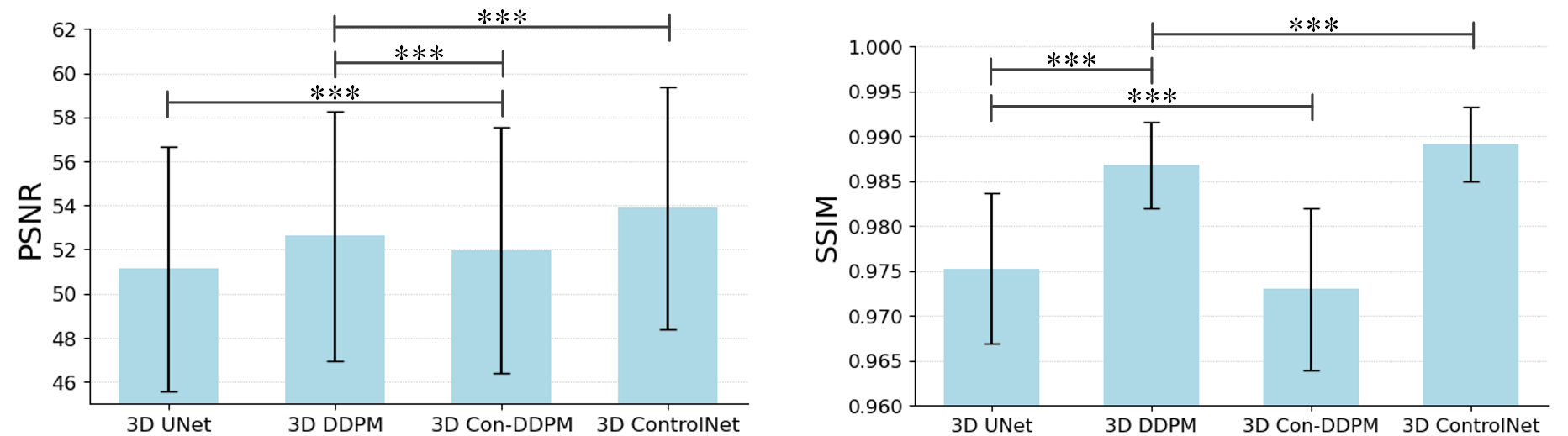}
\caption{The PSNR and SSIM values calculated based on 60 $1/20$ low-dose test datasets. *** located at the top of the bar plot represents p-value \textless $0.001$.}
\label{fig_quant}
\end{figure*}

We trained and evaluated the proposed model using the Siemens Biograph Vision Quadra data from the Ultra-low Dose PET Imaging Challenge, which includes 377 total-body $\rm^{18}F$-FDG PET scans. During the pre-training stage, we utilized 302 normal-dose PET images to train the score function. During the fine-tuning stage, 50 pairs of $1/20$ low-dose and normal-dose PET images were adopted as the low- and high-quality paired data to train the 3D ControlNet-based score function. The validation set consisted of 15 paired PET images, while the test set included 60 paired PET images. Preprocessing steps such as data cropping and unit conversion, as well as the 3D patch overlapping technique used during the inference phase, were implemented following the methodology of Yu et al.~\cite{yu2024pet}.

To evaluate the effectiveness of the proposed method, we compared it with several reference methods. Specifically, to validate the efficacy of fine-tuning based on 3D ControlNet, we directly applied the pre-trained 3D DDPM model for inference and embedded the low-dose PET images as data consistency constraints during the sampling process to obtain the corresponding denoised results, following the approach of Gong et al.~\cite{gong2024pet} and this method was denoted as 3D DDPM. As an additional reference, we trained a conditional 3D DDPM~\cite{yu2024pet} (denoted as 3D Con-DDPM) and a 3D UNet~\cite{cciccek20163d} directly on the same 50 pairs of $1/20$ low-dose and normal-dose PET data. The number of time steps $T$ was set to $1000$ throughout the forward and reverse diffusion processes, and a linear noise schedule was implemented. The training batch size was set to $6$, utilizing $6$ NVIDIA A100 GPUs for distributed training. All methods were implemented using PyTorch, and the learning rate was set to $1 \times 10^{-4}$. Optimizing the 3D DDPM with 3D ControlNet required approximately $24.2\%$ more GPU memory and $27.8\%$ more time per training iteration than optimizing the 3D DDPM without ControlNet.

To quantitatively evaluate the performance of the various methods, we adopted the Peak Signal-to-Noise Ratio (PSNR) and Structural Similarity Index Measure (SSIM), using the normal-dose PET images as the ground truth. Wilcoxon signed-rank tests for PSNR and SSIM were performed for statistical analysis.

\subsection{Results} 

The qualitative results of all methods are shown in Fig.~\ref{fig_visual}. It can be observed that the results produced by the 3D UNet were overly blurry. Although 3D DDPM demonstrated good denoising performance in most regions, noise remained in areas such as the liver and spine. These residual noise artifacts could be mistaken for lesions, potentially leading to serious implications in clinical diagnosis. The 3D Con-DDPM and 3D ControlNet methods generated more realistic denoised results. However, it can be observed that the 3D Con-DDPM suffered from lower contrast and inadequate handling of local details. In contrast, results from the 3D ControlNet were the closest to the ground truth, producing better structural details and more precise edge contours. The quantitative results shown in Fig.~\ref{fig_quant} were consistent with our qualitative observations. The 3D ControlNet achieved better performance in both PSNR and SSIM than the other three methods.

\section{Conclusion}
\label{sec:conclusion}
In this work, we proposed a novel 3D ControlNet-based method for whole-body PET image denoising. By fine-tuning a pre-trained 3D DDPM using a small set of paired low-dose and normal-dose PET images, the proposed approach effectively overcame the limitations of existing diffusion-based denoising methods, which often struggled with adaptability to different acquisition protocols and the risk of overfitting with limited data. The 3D ControlNet architecture preserved the strengths of the large pre-trained model while also incorporating conditional information from low-dose PET images, enabling precise reconstruction of local details essential for clinical diagnosis. Experimental results demonstrated that the proposed method outperformed existing techniques both visually and quantitatively. Our future work will focus on further evaluations based on clinical datasets from various PET protocols.

\section{Compliance with ethical standards}
\label{sec:ethics}
All procedures performed in studies involving human participants were in accordance with the ethical standards of the institutional and/or national research committee and with the 1964 Helsinki declaration and its later amendments or comparable ethical standards. 

\section{Acknowledgments}
\label{sec:acknowledgments}
This work was supported by NIH grants R01EB034692 and R01AG078250.

\bibliographystyle{IEEEbib}
\bibliography{refs}

\begin{thebibliography}{10}

\bibitem{ho2020denoising}
Jonathan Ho, Ajay Jain, and Pieter Abbeel,
\newblock ``Denoising diffusion probabilistic models,''
\newblock {\em Advances in neural information processing systems}, vol. 33, pp. 6840--6851, 2020.

\bibitem{xie2023ddpet}
Huidong Xie, Weijie Gan, Bo~Zhou, Xiongchao Chen, Qiong Liu, Xueqi Guo, Liang Guo, Hongyu An, Ulugbek~S Kamilov, Ge~Wang, et~al.,
\newblock ``Ddpet-3d: Dose-aware diffusion model for 3d ultra low-dose pet imaging,''
\newblock {\em arXiv preprint arXiv:2311.04248}, 2023.

\bibitem{jiang2023pet}
Caiwen Jiang, Yongsheng Pan, Mianxin Liu, Lei Ma, Xiao Zhang, Jiameng Liu, Xiaosong Xiong, and Dinggang Shen,
\newblock ``Pet-diffusion: Unsupervised pet enhancement based on the latent diffusion model,''
\newblock in {\em International Conference on Medical Image Computing and Computer-Assisted Intervention}. Springer, 2023, pp. 3--12.

\bibitem{shen2023pet}
Chenyu Shen, Ziyuan Yang, and Yi~Zhang,
\newblock ``Pet image denoising with score-based diffusion probabilistic models,''
\newblock in {\em International Conference on Medical Image Computing and Computer-Assisted Intervention}. Springer, 2023, pp. 270--278.

\bibitem{pan2024full}
Shaoyan Pan, Elham Abouei, Junbo Peng, Joshua Qian, Jacob~F Wynne, Tonghe Wang, Chih-Wei Chang, Justin Roper, Jonathon~A Nye, Hui Mao, et~al.,
\newblock ``Full-dose whole-body pet synthesis from low-dose pet using high-efficiency denoising diffusion probabilistic model: Pet consistency model,''
\newblock {\em Medical Physics}, 2024.

\bibitem{gong2024pet}
Kuang Gong, Keith Johnson, Georges El~Fakhri, Quanzheng Li, and Tinsu Pan,
\newblock ``Pet image denoising based on denoising diffusion probabilistic model,''
\newblock {\em European Journal of Nuclear Medicine and Molecular Imaging}, vol. 51, no. 2, pp. 358--368, 2024.

\bibitem{xie2024dose}
H~Xie, W~Gan, X~Chen, B~Zhou, Q~Liu, M~Xia, X~Guo, Y-H Liu, H~An, US~Kamilov, et~al.,
\newblock ``Dose-aware diffusion model for 3d low-count cardiac spect image denoising with projection-domain consistency,''
\newblock in {\em 2024 IEEE Nuclear Science Symposium (NSS), Medical Imaging Conference (MIC) and Room Temperature Semiconductor Detector Conference (RTSD)}. IEEE, 2024, pp. 1--1.

\bibitem{zhang2023adding}
Lvmin Zhang, Anyi Rao, and Maneesh Agrawala,
\newblock ``Adding conditional control to text-to-image diffusion models,''
\newblock in {\em Proceedings of the IEEE/CVF International Conference on Computer Vision}, 2023, pp. 3836--3847.

\bibitem{rombach2022high}
Robin Rombach, Andreas Blattmann, Dominik Lorenz, Patrick Esser, and Bj{\"o}rn Ommer,
\newblock ``High-resolution image synthesis with latent diffusion models,''
\newblock in {\em Proceedings of the IEEE/CVF conference on computer vision and pattern recognition}, 2022, pp. 10684--10695.

\bibitem{yu2024pet}
Boxiao Yu, Savas Ozdemir, Yafei Dong, Wei Shao, Kuangyu Shi, and Kuang Gong,
\newblock ``Pet image denoising based on 3d denoising diffusion probabilistic model: Evaluations on total-body datasets,''
\newblock in {\em International Conference on Medical Image Computing and Computer-Assisted Intervention}. Springer, 2024, pp. 541--550.

\bibitem{cciccek20163d}
{\"O}zg{\"u}n {\c{C}}i{\c{c}}ek, Ahmed Abdulkadir, Soeren~S Lienkamp, Thomas Brox, and Olaf Ronneberger,
\newblock ``3d u-net: learning dense volumetric segmentation from sparse annotation,''
\newblock in {\em Medical Image Computing and Computer-Assisted Intervention--MICCAI 2016: 19th International Conference, Athens, Greece, October 17-21, 2016, Proceedings, Part II 19}. Springer, 2016, pp. 424--432.

\end{thebibliography}

\end{document}